# Ultrafast switching of photoinduced phonon chirality in the antiferrochiral BPO$_4$ crystal

Hao Chen[a],[†],[‡] Hanyu Wang[a],[¶],[§] Tingting Wang,[‡] Yongsen Tang,[†] Haoshu Li,[*],[∥] Xiaohong Yan,[*],[†] and Lifa Zhang[*],[‡]

†School of Science, Nanjing University of Posts and Telecommunications, Nanjing 210023, China

‡Phonon Engineering Research Center of Jiangsu Province, Center for Quantum Transport and Thermal Energy Science, Institute of Physics Frontiers and Interdisciplinary Sciences, School of Physics and Technology, Nanjing Normal University, Nanjing 210023, China

¶Key Laboratory of Materials Physics, Institute of Solid State Physics, HFIPS, Chinese Academy of Science, Hefei 230031, China

§Science Island Branch of the Graduate School, University of Science and Technology of China, Hefei 230026, China

∥Department of Physics, University of Science and Technology of China, Hefei 230026, China

E-mail: lihaoshu@ustc.edu.cn; yanxh@njupt.edu.cn; phyzlf@njnu.edu.cn

[a]These authors contributed equally to this work.

**Abstract**

In crystalline systems, chiral crystals cannot interconvert to their enantiomorph post-synthesis without undergoing melting-recrystallization processes. However, recent work indicates that ultrafast terahertz-polarized light has been shown to enable dynamic control of structural chirality in the antiferrochiral boron phosphate (BPO$_4$) crystal. Here, using first-principles calculations and nonlinear phonon dynamics simulations, we investigate the underlying physics of lattice dynamics in this system. The results demonstrate that polarized optical pumping not only induces chiral phonons but also establishes a chirality-selective filtering mechanism, both of which can be reversibly switched by tuning the polarization of the excitation pulse. Furthermore, under a temperature gradient, the pump-induced chiral phonons give rise to ultrafast phonon magnetization, with its direction also controllable via light polarization. Our findings establish a new paradigm for ultrafast optical control of phonon chirality via dynamic chirality switching, offering promising opportunities for chiral information transfer and the design of chiral phononic devices.

Chirality refers to the geometric property of an object that cannot be superimposed onto its mirror image through any combination of rotations or translations and chiral crystals play a fundamental role across chemistry, biology, and physics.[1–3] Their applications are vast, including enantioselective catalysis,[4] chiral drug design,[5] and chiral-induced spin selectivity,[6–8] etc. Owing to their symmetry constraints, chiral crystals provide an exceptional platform for studying chiral phonons.[9] These chiral quasiparticles in chiral crystals have been observed to drive the spin Seebeck effect,[10] generate spin polarization,[11] exhibit selective coupling with photons,[12] produce phonon angular momentum selectivity.[13,14] These findings collectively highlight the significant potential of chiral phonons in quantum manipulation and information transmission within chiral crystals, making it of great interest to exploit the helicity of chiral crystals to propagate these chiral phonons.

Furthermore, achieving control over the directional propagation of specific phonon chirality requires a reliable method for switching structural chirality. However, in conventional chiral materials, the intrinsic structural chirality is typically fixed during crystal growth, making post-synthesis chirality configuration flipping extremely challenging.[15,16] Recently, Zeng et al.[17] reported a breakthrough discovery in the antiferrochiral $BPO_4$ crystal: this unique system consists of alternating structural units with opposite handedness.[18,19] By selectively exciting the system with terahertz light of specific polarization, they demonstrated controlled structural distortions that enable dynamic switching of chirality. This discovery raises a compelling question: Can optical excitation be used to achieve ultrafast switching and selective filtering of chiral phonon propagation in this antiferrochiral system?

Building upon this framework, we investigate the ultrafast optical control of chiral phonon in the antiferrochiral $BPO_4$ crystal under ultrafast optical pumping. Our results demonstrate that pump along different crystal axes not only enables deterministic switching of structural chirality, but also induced directional reversal of phonon chirality in specific vibrational modes, establishing an effective chiral phonon filtering mechanism. Time-resolved simulations allow us to systematically track the evolution of phonon chirality following optical excitation. Furthermore, by characterizing phonon angular momentum under temperature gradients, we reveal ultrafast switching of the net phonon angular momentum and its associated magnetization with alternating optical pump polarization axes. These findings underscore the unique potential of ultrafast optical excitation for dynamic control of chiral phonon transport, offering a new paradigm for active manipulation of phononic chirality and paving the way toward quantum information technologies based on chiral phonons.

***Light-induced chirality in antiferrochiral $BPO_4$ crystals.*** At equilibrium, $BPO_4$ adopts an antiferrochiral crystal structure belonging to the non-chiral space group $I\bar{4}$ (No. 84). Its unit cell consists of two substructural units with opposite handedness, which alternate periodically along the $c$-axis via screw-axis symmetry, as illustrated in Figure 1. First-principles calculations confirm that no chiral phonons exist in this ground-state configuration (see Supporting Information for details). Under terahertz pulse excitation, however, incident light polarized along either the $a$- or $b$-axis can selectively drive specific infrared-active phonon modes. This interaction lifts the degeneracy between the two substructures and induces a transient lattice distortion with well-defined chirality, enabling selective generation of chiral structures.[17]

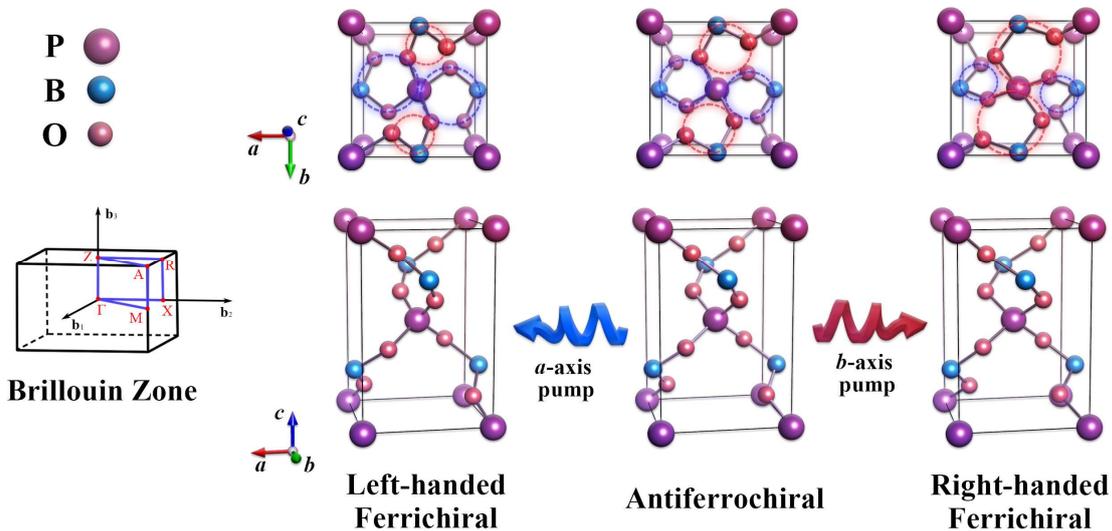

Figure 1: Antiferrochiral and ferrichiral structures of BPO$_4$ and their corresponding Brillouin zone. Left: left-handed ferrichiral structure; center: antiferrochiral structure; right: right-handed ferrichiral structure. Blue and red circles highlight the characteristic of the left- and right-handed substructures, respectively.

We first performed calculations on the antiferrochiral unit cell of BPO$_4$. The optimized lattice constants were found to be $a = b = 4.34\,\text{Å}$ and $c = 6.65\,\text{Å}$, consistent with previously reported values.[20] Based on this relaxed structure, we applied nonlinear phonon dynamics theory[21–24] to investigate transient structural distortions induced by optical polarized excitation along the $a$- and $b$-axes. In BPO$_4$, two orthogonal and degenerate infrared-active phonon modes, $Q_{E,a}$ and $Q_{E,b}$, can be selectively driven by the polarized light. Through the lowest-order nonlinear coupling terms:[17]

$$U = -\alpha Q_{E,a}^2 Q_B + \alpha Q_{E,b}^2 Q_B, \tag{1}$$

excitation of the E-symmetry phonon modes induce a displacement force on the B-symmetry chiral phonon mode ($Q_B$), effectively enabling chirality control. Specifically, an $a$-axis-polarized terahertz pulse exciting $Q_{E,a}$ leads to a positive displacement of $Q_B$, driving the system into a left-handed ferrichiral structure. In contrast, excitation along the $b$-axis induces a right-handed ferrichiral distortion. These two photoinduced ferrochiral structures, shown in the left and right panels of Figure 1, retain the tetragonal lattice character but exhibit significantly reduced symmetry, belonging to the space group $P1$ (No. 1).

*Light-induced ultrafast switching of chiral phonons*. The two ferrichiral structures of BPO$_4$ induced by optical pumping both belong to the chiral, non-centrosymmetric space group $P1$, which breaks the original inversion symmetry ($I$) of the antiferrochiral phase—thus fulfilling the fundamental symmetry requirement for the emergence of chiral phonons. Using first-principles calculations, we systematically analyzed the phonon polarization $s_{\mathrm{ph}}^z$ characteristics of these two ferrichiral structures along several high-symmetry paths in the Brillouin zone. The definition of phonon polarization is shown in the Supporting Information. Notably, some phonons along the $\Gamma - Z$ direction (parallel to the $c$-axis) exhibit pronounced chirality. The corresponding phonon dispersion and polarization along this path are shown in Figure 2 (a) and (b), which show the phonon polarization distributions at 0.2 ps after optical excitation using $a$- and $b$-axis polarized light with an excitation fluence, respectively. Red branches represent right-handed phonons with positive phonon polarization, while blue branches indicate left-handed phonons with negative polarization. It is worth noting that these ferrichiral structures are transient, non-equilibrium states induced by optical excitation; as a result, some imaginary frequencies appear in the calculated phonon dispersion. In comparison to the antiferrochiral state, several previously degenerate modes undergo noticeable splitting, with the resulting doublet modes displaying opposite polarization signs. Notably, along the $\Gamma - Z$ path, phonons in the same ferrichiral structure exhibit strict antisymmetric polarization with respect to positive and negative wave vectors, while phonons with the same wave vector direction in opposite ferrichiral domains exhibit reversed polarization signs. The atomic displacement patterns and polarization directions of representative modes are also provided in the Supporting Information.

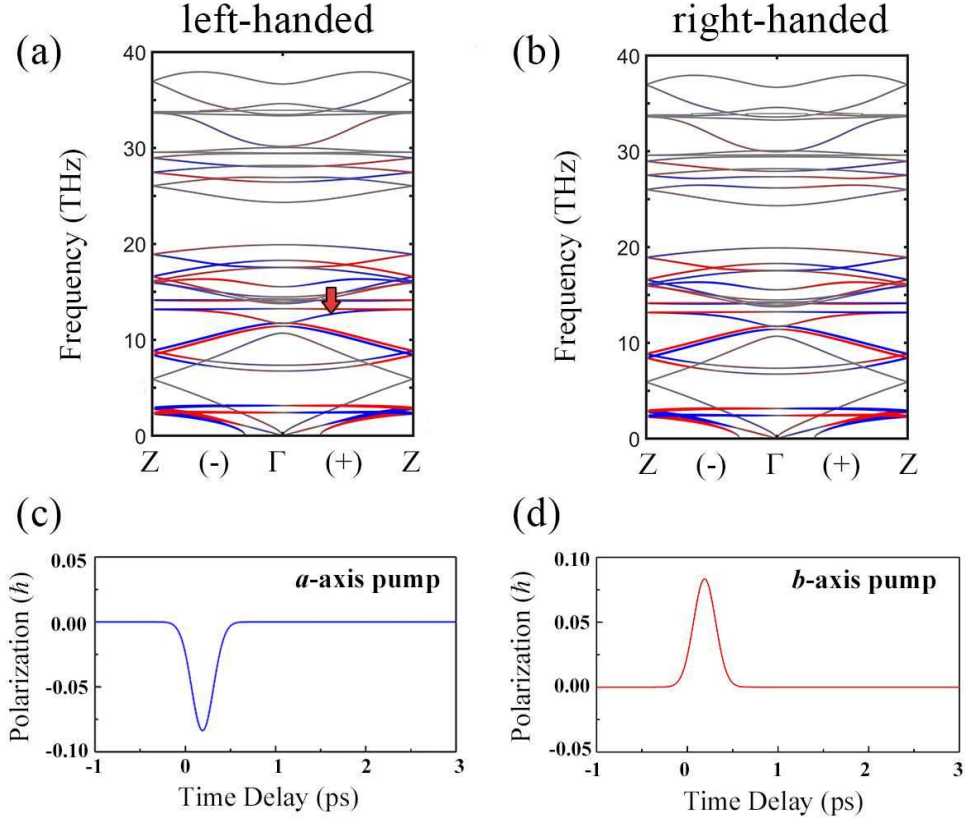

Figure 2: Phonon polarization of $BPO_4$ ferrichiral structures under optical pumping. (a) Left-handed ferrichiral structure and (b) right-handed ferrichiral structure: phonon polarization dispersions along the high-symmetry $\Gamma - Z$ path at 0.2 ps after excitation. The (+) and (−) signs along the path denote phonon wave vectors with $k_z > 0$ and $k_z < 0$, respectively. Red and blue branches in the dispersions correspond to positive and negative phonon polarization values, with the linewidth indicating the magnitude of the polarization. (c) Left-handed ferrichiral structure and (d) right-handed ferrichiral structure: time evolution of the phonon polarization for the 11th vibrational mode at the midpoint of the positive $\Gamma - Z$ direction following optical excitation.

To further investigate the ultrafast dynamics of chiral phonons, we selected the 11th phonon mode (highlighted by the red arrow in Figure 2) as a representative example. We simulated the time evolution of its polarization at the midpoint of the positive $\Gamma-Z$ direction under excitation by differently polarized pump light, as shown in Figure 2 (c) and (d). The results show that the phonon polarization rapidly reaches a peak of 0.08 $\hbar$ within 0.2 ps after excitation and decays to zero within approximately 0.5 ps, closely tracking the ferrichiral structural transition. Importantly, the polarization induced by pump light along orthogonal directions has equal magnitude but opposite sign, consistent with the phonon polarization dispersion analysis. These results demonstrate that by controlling the pump-light polarization direction, ultrafast switching of phonon chirality can be achieved on the picosecond timescale through photoinduced ferrichiral structural distortions.

*Light-switchable chirality filtering effect.* Within a specific frequency range, we observe a strict correspondence between phonon chirality and structural handedness for forward-propagating phonons (i.e., modes with positive group velocity) in the left- and right-handed ferrichiral structures. As illustrated by the 11th phonon branch in Figure 2 (a), all forward-propagating phonons in the left-handed ferrichiral structure exhibit negative polarization, indicating purely left-handed character. This implies that only left-handed phonons are allowed to propagate in this direction. In contrast, the right-handed ferrichiral structure supports only right-handed phonons for unidirectional transport. This chirality-dependent propagation behavior provides a crucial theoretical foundation for realizing the chiral phonon filtering effect in $BPO_4$.

To illustrate the chiral phonon filtering effect in $BPO_4$, we present a schematic of the physical mechanism in Figure 3. Taking the 11th phonon branch as an example, chosen for its strong polarization and spectral isolation, we demonstrate how selective chiral transport can be realized. As shown in the left panel of Figure 3, when a polarized pump pulse at 19 THz along the *a*-axis induces the formation of the left-handed ferrichiral structure, a second linearly polarized probe beam with a frequency of 12.5 THz (resonant with the 11th branch of phonon) is incident along the $c$-axis from the bottom of the crystal. In the left-handed ferrichiral structure, only left-handed chiral phonons possess positive group velocity at this frequency, allowing them to propagate along the positive $c$-direction while completely suppressing the right-handed components (which have negative group velocity). This selective transport effectively filters out the right-handed information from the incident light, resulting in a pronounced circular dichroism signal at the output. In contrast, the right-handed ferrichiral structure exhibits the opposite filtering behavior, allowing only right-handed phonons to transmit while filtering out the left-handed ones under identical excitation conditions.

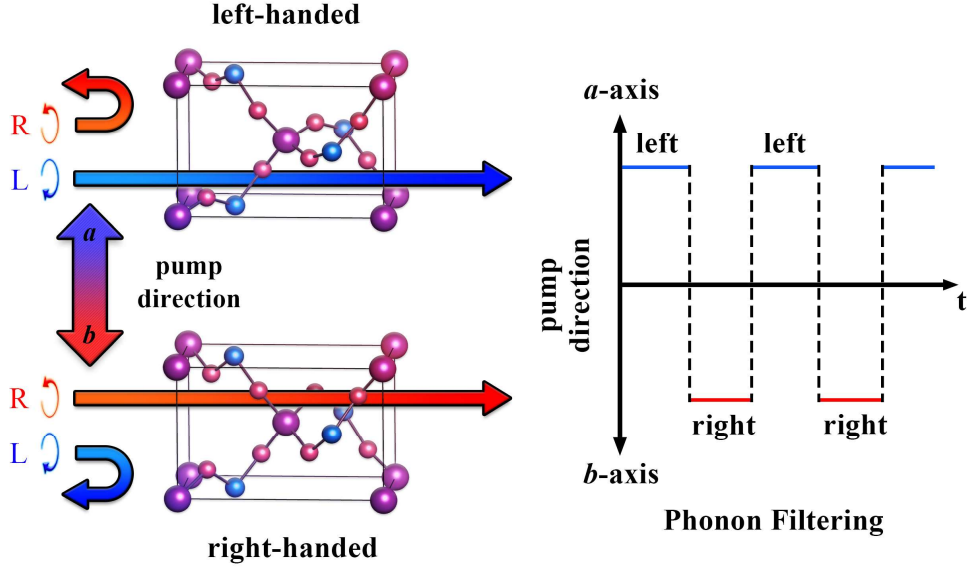

Figure 3: Schematic of the switchable chiral phonon filtering effect in ferrichiral BPO$_4$. Left: horizontal blue and red arrows represent the unidirectional propagation of left- and right-handed phonons in specific ferrichiral structures. Right: illustration of changing the pump light polarization on an ultrafast timescale reverses the chirality of forward-propagating phonons.

Crucially, this chiral filtering functionality can be dynamically reversed by switching the polarization direction of the first pump pulse between the $a$- and $b$-axes, as schematically illustrated in the right panel of Figure 3. The entire process operates on a picosecond timescale, highlighting the potential of ultrafast optical control of phonon chirality for the development of next-generation quantum information technologies.

***Ultrafast light-induced phonon magnetization.*** In thermal equilibrium, nonmagnetic crystals exhibit zero net phonon angular momentum, as required by time-reversal symmetry. This symmetry ensures that the phonon polarization at wavevectors $\pm k$ cancels out. However, when the system is driven out of equilibrium, such symmetry constraints can be lifted. In particular, under an applied temperature gradient $\nabla T$, the phonon distribution function can be expressed as: $f(\boldsymbol{k}, \sigma) = f_0(\boldsymbol{k}, \sigma) - \tau v^i(\boldsymbol{k}, \sigma)\frac{\partial f_0}{\partial T}\frac{\partial T}{\partial D^i}$. The second term on the right side represents that the deviation of the phonon distribution from the Bose distribution $f_0$, where $\tau$ denotes the phonon lifetime, $v^i(\boldsymbol{k}, \sigma)$ is the group velocity of a phonon mode along the $i = x, y, z$ direction, $\boldsymbol{k}$ is the wavevector, $\sigma$ is the phonon branch index and $D^i$ denotes the position component along the $i$ direction. Under a temperature gradient, the population imbalance between phonons of opposite chirality leads to a nonzero net phonon angular momentum, given by:[25]

$$J_{\mathrm{ph}}^i = -\frac{\tau}{V}\sum_{\boldsymbol{k},\sigma} s^i(\boldsymbol{k}, \sigma) v^j(\boldsymbol{k}, \sigma)\frac{\partial f_0}{\partial T}\frac{\partial T}{\partial D^j} \equiv \beta^{ij}\frac{\partial T}{\partial D^j}. \qquad (2)$$

$V$ represents the unit cell volume, and $\beta^{ij}$ is the corresponding response tensor.

Based on the nonequilibrium phonon angular momentum theory described above, we investigated the phonon magnetization behavior of BPO$_4$ under an applied temperature gradient. As illustrated in Figure 4, when a longitudinal temperature gradient $\nabla T$ is applied along the $c$-axis, the dynamically induced structural chirality—triggered by $a$- or $b$-axis polarized optical pumping—generates a transient net phonon angular momentum $J_{ph}$. Our calculations show that, under a positive temperature gradient along the $+c$-axis, the left-handed ferrichiral structure also produces a net $J_{ph}$ pointing along $+c$, whereas switching the pump polarization to induce the right-handed ferrichiral structure reverses the direction of $J_{ph}$ to $-c$. This reveals an ultrafast and reversible modulation of phonon angular momentum via optical pumping control. Furthermore, we performed quantitative calculations of the phonon angular momentum and its associated magnetization. Based on symmetry analysis of the $P1$ space group, we got the magnitude of corresponding response tensor components $\beta^{zz}$ is $10^{-6} \times [\tau/(1\mathrm{s})]\mathrm{J\,s\,m^{-2}\,K^{-1}}$ at room temperature ($T = 300$ K). Since phonon angular momentum originates from the collective rotational motion of positively charged atomic nuclei, this motion directly gives rise to a magnetization effect. Based on the Born effective charge, we can estimate it. The relationship between the angular momentum $\boldsymbol{j}$ and magnetization $\boldsymbol{m}$ is given by $\boldsymbol{m} = \gamma \boldsymbol{j}$, where $\gamma$ is the gyromagnetic ratio. For each atom, the gyromagnetic ratio can be expressed as $\gamma_{\alpha\beta}^{\mathrm{atom}} = geZ^*_{\alpha\beta}/2m_{\mathrm{atom}}$, where $Z^*_{\alpha\beta}$ is the Born effective charge tensor and was calculated from first-principles calculations; $m_{\mathrm{atom}}$ is the mass of different atoms. Assuming that the $g$ factor is in the range of 1-10 and considering a phonon relaxation time $\tau \sim 1$ ps, the magnetization can be estimated as:

$$M_{ph} \sim -\frac{\Delta T/(1\mathrm{K})}{L/(1\mathrm{m})} \times 10^{-10} \mathrm{A/m}. \tag{3}$$

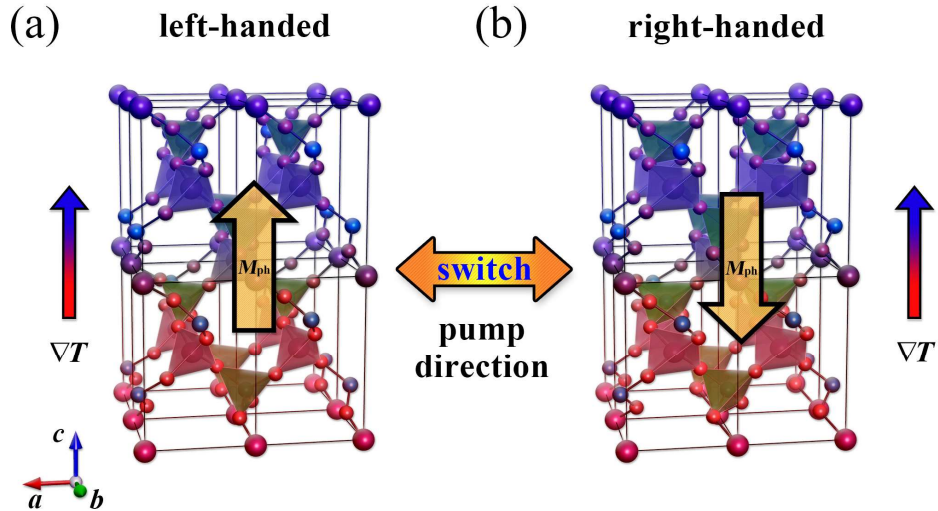

Figure 4: Schematic illustration of the reversal of ultrafast light-induced phonon magnetization ($M_{ph}$) under a longitudinal temperature gradient ($\nabla T$) in different ferrichiral structures.

Over the sample size $L$, the temperature difference is $\Delta T$.

These results reveal that ultrafast light-induced magnetization can emerge in ferrichi-ral $BPO_4$ under a temperature gradient, a process that serves as the opposite to ultrafast demagnetization observed in magnetic systems.[26] Moreover, the direction of the ultrafast light-induced phonon magnetization can be dynamically reversed by switching the polarization axis of the optical pump.

***Discussion and Summary.*** In this work, we investigate light-induced phonon chirality and its ultrafast control mechanisms in the antiferrochiral crystal $BPO_4$. Our results reveal that the phonon chirality of the material can be dynamically and reversibly switched by tuning the polarization direction of an ultrafast pump pulse. This further realizes the selective generation and directional propagation of chiral phonons, giving rise to a robust phonon chirality filtering effect. The light-induced chirality switching operates on an ultrafast timescale with high selectivity, offering a novel pathway for controlling and utilizing phononic chiral information. Furthermore, the emergence of photoinduced phonon magnetization under temperature gradient reveals the connection between dynamic phonon chirality and magnetization, providing new opportunities for the application of ultrafast phonon chirality switching in quantum and spin electronics.

# Supporting Information Available

The first-principles calculations details, phonon chirality and polarization, phonon polarization in antiferrochiral $BPO_4$, atomic circular vibration.

# Acknowledgement

This work was supported by National Key R&D Program of China (2023YFA1407001), Department of Science and Technology of Jiangsu Province (BK20220032) and National Natural Science Foundation of China (12304124). Hao Chen acknowledges support from the Natural Science Research Start-up Foundation of Recruiting Talents of Nanjing University of Posts and Telecommunications (Grant No.NY224105).